\documentclass[astrosymb, twocolumn]{aastex7}
\makeatletter
\let\frontmatter@title@above=\relax
\makeatother

\usepackage{graphicx}
\usepackage{dcolumn}
\usepackage{bm}
\usepackage{amsmath}
\usepackage{amssymb}
\usepackage{mathbbol}
\usepackage[utf8]{inputenc}
\usepackage{hyperref}
\hypersetup{linkcolor=red,citecolor=blue,filecolor=cyan,urlcolor=magenta}

\newcommand       \ba           {\begin{eqnarray*}}
\newcommand       \ea           {\end{eqnarray*}}
\newcommand       \baa           {\begin{equation}}
\newcommand       \eaa           {\end{equation}}

\begin{document}

\title{Latent thermal instability }

\author[orcid=0000-0002-3883-4276,sname='Choudhury']{Prakriti Pal Choudhury}
\altaffiliation{}
\affiliation{Department of Physics, University of Oxford}
\email[show]{prakriti.palchoudhury@physics.ox.ac.uk}  

\author[]{Archie F. A. Bott}
\altaffiliation{}
\affiliation{Department of Physics, University of Oxford}
\email[show]{}

\begin{abstract}
Multiscale temperature fluctuations are abundant in the intracluster medium (ICM) outside of galaxy cluster cores ($\sim 100~{\rm kpc}$). Their origin is often attributed to turbulent stirring by subhalos or accreting baryons crossing the virial radius. However, their apparent resistance to mixing and thermal conduction in a collisional medium has not been explained. We propose a new mechanism by which steady-state temperature fluctuations can form and persist outside the cluster core. Local thermal instability, or Field instability, is used to explain filamentary condensates in cluster cores but is usually dismissed outside them because thermal conduction should suppress instability. In weakly collisional or collisionless plasmas, however, thermal conduction can be anomalously suppressed by  heat-flux-driven plasma instabilities triggered in presence of a local magnetic field, leading to two effects: (i) condensates form in a new parameter regime that overlaps with conditions outside the core, and (ii) condensates reach a steady state as in the hydrodynamic limit. This extends the regime of instability-driven fluctuations to over $\gtrsim50\%$ (depending on hot plasma temperature) of the cluster. We use one-dimensional hydrodynamic simulations of condensates to test our analytical ideas.


\end{abstract}



\section{Introduction}
The regions outside the cores of galaxy clusters have become a new frontier for probing the intracluster medium (ICM), the intergalactic medium, and cosmology, enabled by multi-wavelength synergy between X-ray, Sunyaev-Zeldovich (SZ), and Weak Lensing (WL) observations (see \citealt{walker2019physics} for a review). Cluster thermodynamics is jointly constrained by X-ray and SZ observations, while WL can reveal hydrostatic mass bias. A prime example is the Coma cluster, for which joint observations have revealed pressure fluctuations (e.g., \citealt{2004A&A_schuecker, 2016MNRAS_khatri}) and density fluctuations at large distances from the core (e.g., \citealt{2020MNRAS_mirakhor, 2012MNRAS_churazov}). Generally, temperature contrasts outside the cores, on both small and large scales, have been discussed for several decades (e.g., \citealt{1995A&A_briel, 1999A&A_kull, 1980MNRAS_cowie, 2004ApJ_sparks, 2006ApJ_ascasibar}).

The origins of fluctuations in the ICM outskirts remain debatable, but multiscale temperature, density, and pressure fluctuations perhaps arise from turbulent stirring by accreting plasma or galaxies crossing the virial radius. The puzzling observation is that they do not get erased. For a temperature fluctuation on scale $\ell$, decay should occur at a rate $\kappa_{\rm c} T_0/p_0\ell^2$ in a strongly collisional plasma, where $\kappa_{\rm c}$ is the collisional thermal conductivity (\citealt{1962pfig.book_spitzer}), $T_0$ the mean temperature, and $p_0$ the mean pressure of the medium. This timescale is short outside cluster cores compared to merger-duration timescales, yet fluctuations persist (\citealt{2003ApJ_markevitch}). Temperature structures created on a scale $\ell$ should also de-cohere on a timescale $\ell/u_{\ell}$ (where $u_{\ell}$ is the velocity of the turbulent eddy of size $\ell$) in turbulent media; in other words, once these structures have formed, they will be mixed quickly unless persistently stirred. This raises various questions: why do they survive, and is there an alternative mechanism for producing steady-state multiscale temperature fluctuations? In this work, we propose a new solution to this problem.

Thermal instability (TI), driven by radiative cooling in a dilute magnetized plasma, has been widely discussed as an explanation for density fluctuations or condensates, from solar prominences (e.g., \citealt{1953ApJ_parker}) to the intracluster and intergalactic medium on large scales (e.g., \citealt{1965ApJ_field}, subsequently followed up in simulations of globally thermally stable clusters by \citealt{2012MNRAS_mccourt, 2012MNRAS_sharma, 2019MNRAS_choudhury} and for a mini-review see \citealt{choudhury}). In unmagnetized plasma, thermal conduction sets the smallest scale for thermally unstable modes, the Field length $\lambda_{\rm F}$, below which heat flux dominates over radiative cooling. The linear TI growth rate is scale free. But at large scales, TI transitions from isobaric ($\delta p/p_0 \ll \delta \rho/\rho_0$) to isochoric ($\delta p/p_0 \gg \delta \rho/\rho_0$) because the sound-crossing time exceeds the radiative cooling time, a point emphasized in recent work on condensation in circumgalactic plasma in smaller dark matter halos (e.g., \citealt{Waters2019Non-isobaricInstability, Das2021ShatterInstability}).


The known effect of magnetic field on TI is twofold: it can affect dynamics directly, enhancing condensation  (e.g., \citealt{2018MNRAS_ji, 2019A&A_claes, 2025MNRAS_choudhury, 2025MNRAS_kaul, 2025MNRAS_wibking, 2026arXiv_voit}), and it also modifies the nature of thermal conduction (e.g., \citealt{Sharma2010ThermalClusters}). In this work, we focus on the effect on conduction. For a collisional plasma, \citealt{1965RvPP_braginskii} showed that heat transport is anisotropic when $r_e \ll \lambda_{\rm mfp}\ll \ell_{\rm T,\parallel}$, where $r_e$ is the electron gyroradius, $\lambda_{\rm mfp}$ is the Coulomb collisional mean free path, and $\ell_{\rm T,\parallel}$ is the temperature-gradient scale along the field (which may in principle differ from the size of temperature fluctuations). This is because electrons are constrained to gyrate about the local magnetic field. A tangled magnetic field therefore channels heat along field lines and suppresses the effective isotropic heat flux by a factor of $\sim 3-5$ (\citealt{2001ApJ_narayan}), a point discussed in the context of cooling flows in galaxy clusters (\citealt{1994ARA&A_fabian, 2003ApJ_zakamska, 2009ApJ_parrish}). For TI, this implies that a local patch of tangled field may support fine-structured condensates, but overall the turbulent ICM behaves like a hydrodynamic medium. The key problem therefore remains: classical anisotropic conduction is expected to erase TI-driven temperature contrasts on relatively short timescales, especially far from cluster centers, since the radiative cooling time increases with radius (see \citealt{2004MNRAS_nipoti} for constraints on filament sizes across background temperatures).

However, this conclusion overlooks the anomalous transport properties of weakly collisional and collisionless plasmas. Since density decreases with radius in galaxy clusters, the collisional mean free path becomes larger than in the core, so $\lambda_{\rm mfp}\lesssim \ell_{\rm T}$. In a collisionless medium, non-local effects can regulate heat transport because fast electrons stream along magnetic fields (\citealt{Cowie1977THERATES}). In addition, if magnetization is weak, microscale electromagnetic plasma instabilities also anomalously regulate transport~(\citealt{bott_kinetic_2024}). The condition for the heat-flux-driven whistler instability, a kinetic electromagnetic microinstability that grows at electron Larmor scales (\citealt{levinson1992inhibition, Pistinner1998}), is $\beta_e \lambda_{\rm mfp}/\ell_{\rm T}\gtrsim 1$, where $\beta_e=8\pi p_0/B^2_0$ is the ratio of thermal pressure ($p_0$) to magnetic pressure ($B^2_0/8 \pi$), and this condition is often satisfied in the ICM. Particle-in-cell simulations across a range of $\ell_{\rm T}$ and numerical setups robustly show that this instability suppresses the free-streaming heat flux by a factor proportional to $\beta_e$ (\citealt{Roberg-Clark2018SuppressionGradient, 2018JPlPh_komarov, Yerger2024CollisionlessTurbulence, 2025arXiv_choudhury,lopez_collisional_2025}). The whistler waves scatter electrons in pitch angle, limiting transport to the whistler phase speed. Recent laser-plasma experiments have also hinted at suppressed thermal conduction in high-$\beta_e$ plasmas (\citealt{meinecke_strong_2022,vincent_design_2026}). In galaxy clusters, $\beta_e \lambda_{\rm mfp}/\ell_{\rm T}$ spans a wide range, so classical transport models may not provide the most accurate description of thermal conduction everywhere. 

Some effort has already been made to assess the role of anomalous transport in modifying fluid instabilities (see \citealt{2024A&A_perrone, 2022A&A_beckmann}); here, we examine for the first time the effect of anomalous conduction on TI.  
We identify a condensation regime below the classical Field length, which we call Latent TI (LaTI). We develop one-dimensional (1D) hydrodynamic simulations, assuming cylindrical hot plasma filaments aligned with a coherent magnetic field, parameterized by the electron plasma $\beta_e$ and the perturbation size corresponding to the initial coherence scale $\propto k_0^{-1}$, with $k_0$ the wave number, to validate the LaTI regime and explore basic condensate properties when the filament grows clumps along itself.

In section \ref{sec: prelim}, we describe the theoretical preliminaries and physical assumptions. In section \ref{sec:sim}, we discuss the 1D hydrodynamic simulations and  LaTI properties. In section \ref{sec:cluster_disc}, we apply LaTI to the galaxy-cluster context, and in section \ref{sec:disc}, we conclude.

\section{Theoretical preliminaries}
\label{sec: prelim}
\begin{figure}
    \includegraphics[width=9cm]{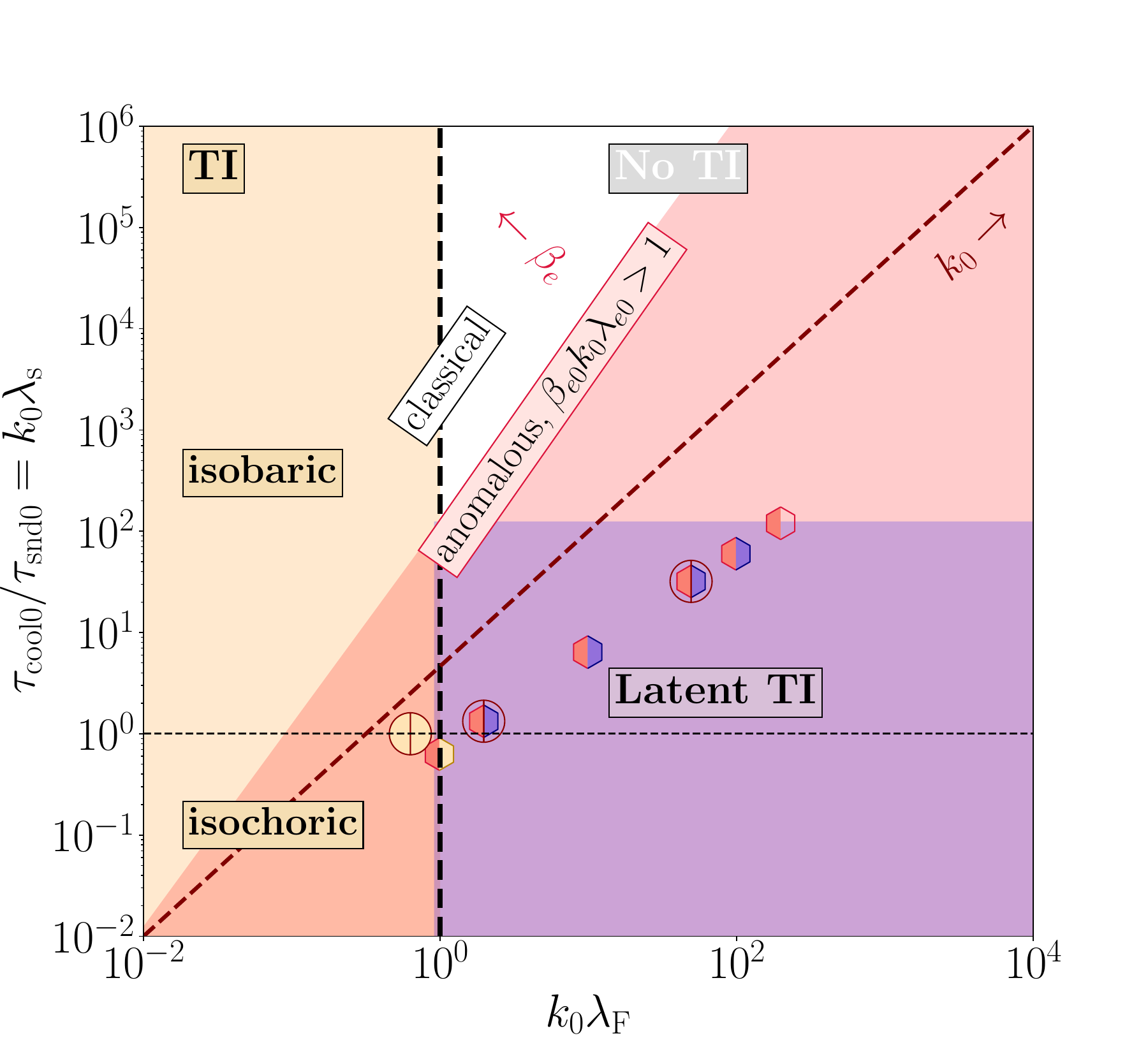}
    \centering
    \caption{A parametric space that defines the latent thermal instability (LaTI) zone in terms of the ratio of cooling time $\tau_{\rm cool0}$ to sound-crossing time $\tau_{\rm snd0}$ on the $y$-axis and the normalised wave number corresponding to the initial temperature gradient scale ($k_0=2\pi/\ell_{\rm T,\parallel0}$) with respect to the classical Field length on the $x$-axis. The $y$-axis is also equivalent to the gradient scale with respect to the length $\lambda_{\rm s}$ that sound crosses in a cooling time. The yellow zone is the TI region for classical thermal conduction, the red zone {\color{black} (to the right of the line marked by $\beta_{e0}k_0\lambda_{e0}=1$)} is the anomalous conduction region, and the purple zone, {\color{black}inside the anomalous conduction region,} is the LaTI region which opens up thermal instability at small scales ($k_0\lambda_{\rm F}\gg1$). {\color{black} We show the parameter space for $\beta_{e0}=5000$ and indicate the expansion of the red region for higher $\beta_{e0}$ with a red arrow.} The hexagons show simulations including anomalous physics and the circles show simulations with only classical physics. The colors on the markers indicate the dominant physical processes - red-purple hexagons with anomalous conduction and LaTI, red-yellow hexagon with dominant TI entering the previously known regime, half red in the anomalous region but without LaTI. }
    \label{fig:fig1}
\end{figure}
We consider a 1D medium in the $\hat{\boldsymbol{x}}$ direction, aligned with the local magnetic field and thus cylindrically symmetric, with approximately uniform number density ($n=n_e\sim n_i$) and temperature ($T=T_i\sim T_e$), where the subscripts $e,i$ denote electrons and ions, respectively. Electron-ion collisions cause electrons to accelerate and emit radiation at a rate proportional to $T_e^{{1}/{2}}$ via Bremsstrahlung cooling, such that the rates of ionization and recombination are balanced. We assume that the medium is heated on average at the rate $\mathcal{H}=\langle \zeta \rangle$, where $\zeta$ is the radiative cooling rate. This heat is supplied by energy sources such as AGNs, supernovae, or mergers, the details of which are ignored for the present purpose\footnote{It is not entirely clear how the hot atmospheres in clusters are maintained in overall thermal equilibrium, but the assumption is justified by the energy budget from the aforementioned heat sources.}. The medium is collisional and follows mass, momentum, and energy conservation according to hydrodynamic evolution ($\lambda_{\rm mfp}\ll \ell_{{\rm T}}$). The equations relevant to such an evolution are the following:
\begin{eqnarray}
    \label{eq:hyd1}
\frac{D \rho}{Dt} &=& -\rho \boldsymbol{\nabla}\cdot \boldsymbol{v},\\
\label{eq:hyd2}
\frac{D \boldsymbol{v}}{Dt} &=& -\frac{\boldsymbol{\nabla}p}{\rho}, \\
\label{eq:hyd3}
\frac{p}{\gamma - 1} \frac{D \ln \left({p}/{\rho^{\gamma}}\right)}{Dt} &=& -\mathcal{\zeta} +\mathcal{H} + \boldsymbol\nabla \cdot \left({\kappa_{\rm (c,a),\parallel} \boldsymbol{\nabla} T}\right),
\end{eqnarray}
\\
where $\frac{D}{Dt}= \frac{\partial}{\partial t} + \boldsymbol{v}\cdot \boldsymbol{\nabla}$ is the Lagrangian derivative, $p$, $\rho$, $T$, and $\boldsymbol{v}$ are the pressure, density, temperature, and velocity (which is approximately the bulk velocity $\boldsymbol{v}_i$ of the ions) of the fluid, and $\gamma=5/3$ is the adiabatic index of an ideal gas. In eq. \ref{eq:hyd3}, the last term is related to the parallel heat flux $\mathcal{Q}$, given by
\begin{equation}
    \mathcal{Q} = \kappa_{\rm (c,a), \parallel} \boldsymbol{\nabla_{\|}} T.
\end{equation}
\begin{figure*}
    \includegraphics[width=18cm]{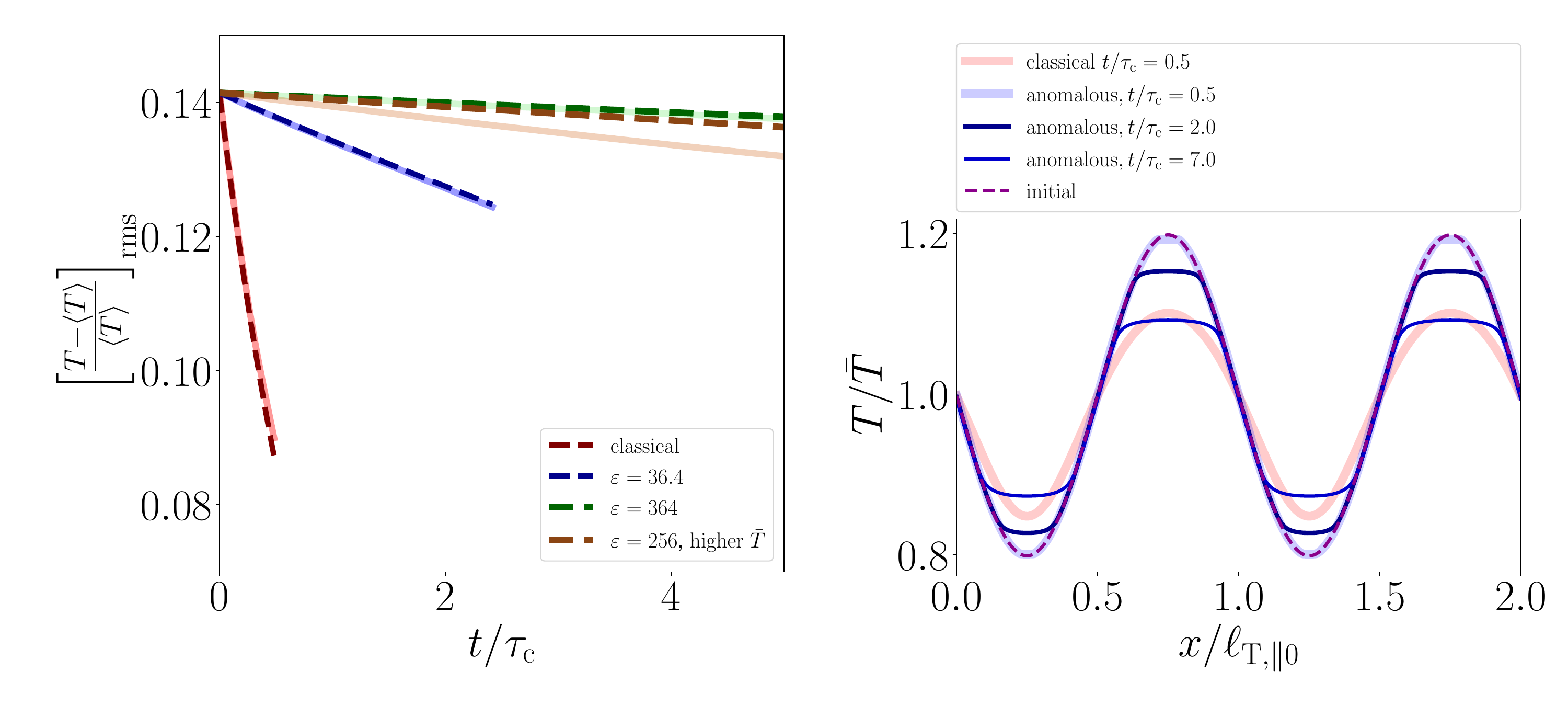}
    \centering
    \caption{{\it Left:} The decay of rms temperature fluctuation (with anomalous suppression and no radiative cooling) with time normalised by classical conduction rate. Solid lines represent simulations and dashed lines show exponential decay. Different colors demonstrate different $\varepsilon= \beta_{e0}\lambda_{e0}/\ell_{\rm T,\parallel0}$. {\it Right:} Comparison of the temperature profiles (normalised to initial mean temperature) for anomalous (blue) and classical (red) with $\varepsilon=36.4$.}
    \label{fig:fig2}
\end{figure*}

For a collisional plasma, parallel heat transport is governed by collisions of free electrons with background electrons and ions parallel to the local magnetic field, and so the parallel thermal conductivity in this case is,
\begin{equation}
\label{eq: clascond}
\kappa_{\rm c,\parallel} = 2.6\times 10^{11} \left({\frac{T}{\rm keV}}\right)^{\frac{5}{2}}~{\rm erg~{cm}^2~s^{-1}{K}^{-1}}.
\end{equation}
If a local patch of this medium evolves such that $\lambda_e \beta_e\gg \ell_{{\rm T},\parallel}$, kinetic instabilities are expected to suppress the heat flux (free-streaming electrons) down a temperature gradient by a factor $\beta_e^{-1}$. Since the medium around galaxies has a large $\beta_e$, this modification to transport may occur in such media, particularly at large radii away from the central galaxy. A proposed sub-grid closure for parallel thermal conductivity in such a scenario is (\citealt{2018JPlPh_komarov}),
\begin{equation}
\label{eq: anocond}
\kappa_{\rm a, \parallel} = \frac{2.6\times 10^{11}}{1+ \mathcal{A}\lambda_e \beta_e/\ell_{{\rm T},\parallel} } \left({\frac{T}{\rm keV}}\right)^{\frac{5}{2}}~{\rm erg~{cm}^2~s^{-1}{K}^{-1}},
\end{equation}
where we used $\mathcal{A}=0.53$ and $\ell_{{\rm T},\parallel}={|\boldsymbol{\nabla}_{\parallel}\ln T|}^{-1}$, the scale of the temperature gradient along the local magnetic field. Thermal conductivity thus smoothly interpolates between the classical and anomalous regimes. In our current work, we consider a plasma at fixed plasma $\beta_e$. We also use the following for the collisional mean free path:
\begin{equation}
\label{eq:mfp}
\lambda_e = 0.003 {\left( \frac{T}{{\rm keV}} \right)}^2{\left(\frac{n}{0.1~{\rm cm}^{-3}}\right)}^{-1}~{\rm kpc}.
\end{equation}
The RHS of eq. \ref{eq:hyd3} in exact balance ($\mathcal{H}=0$) defines the Field length for classical conduction, {\color{black}which we define as $\lambda_{\rm F} = {[\kappa_{\rm c,\parallel} \tau_{\rm cool}(\gamma -1)/(nk_{\rm B})]}^{1/2}$, in other words}
\begin{equation}
\label{eq:fldlen}
\lambda_{\rm F}=9.63 {\left( \frac{T}{{\rm keV}} \right)}^{5/4}{\left( \frac{\tau_{\rm cool}}{1~{\rm Gyr}}\right)}^{1/2}{\left(\frac{n}{0.1~{\rm cm}^{-3}}\right)}^{-1/2}~{\rm kpc} ,
\end{equation}
where the cooling time is {\color{black} $\tau_{\rm cool}=nk_{\rm B}T/[(\gamma-1)\zeta]$ and $\zeta\propto n^2T^{1/2}$ for free-free emission. This can be written as,}
\begin{equation}
    \label{eq:coolt}
    \tau_{\rm cool} = 0.09 {\left(\frac{T}{{\rm keV}}\right)}^{1/2}{\left(\frac{n}{0.1~{\rm cm}^{-3}}\right)}^{-1}~{\rm Gyr}.
\end{equation}
Below this scale, conduction is expected to erase temperature fluctuations. The decay timescale for a given length scale $\ell$ due to classical conduction is {\color{black} $\tau_{\rm c}=\ell^2n k_{\rm B}/[\kappa_{\rm c, \parallel}(\gamma-1)]$, or}
\begin{equation}
    \label{eq:condt}
    \tau_{\rm c} =  0.03 {\left(\frac{\ell}{1~{\rm kpc}}\right)}^2{\left( \frac{T}{{\rm keV}}\right)}^{-5/2} \left(\frac{n}{0.1~{\rm cm^{-3}}}\right) ~{\rm Gyr}.
\end{equation}
We also frequently consider the characteristic length scale that sound crosses in a cooling time, 
\begin{equation}
    \label{eq:lams}
    \lambda_{\rm s} = c_{\rm s}\tau_{\rm cool}=45 \left( \frac{T}{{\rm keV}} \right)\left(\frac{n}{0.1~{\rm cm^{-3}}}\right)^{-1}~{\rm kpc} , 
\end{equation}
and the associated sound-crossing timescale over a length $\ell$:
\begin{eqnarray}
    \label{eq: sndt}
    &\tau_{\rm snd}& = {\color{black}\frac{\ell}{c_{\rm s}}}= 0.002\left(\frac{\ell}{1~{\rm kpc}}\right) {\left( \frac{T}{{\rm keV}}\right)}^{-1/2}~{\rm Gyr}.
\end{eqnarray}
\subsection{What happens to the TI criterion?}
\label{sec:conds}
In terms of microphysical quantities, the classical parallel conductivity scales as $\kappa_{\rm c,\parallel}\sim n_e \lambda_e v_{{\rm th},e}$. It follows that TI growth is suppressed by conduction under the following condition on the rates of the two processes at a wavenumber $k_{\ell}$ associated with the length scale $\ell$:
\begin{equation}
\label{eq:clasField}
{k^2_{\ell} \lambda_{e0} v_{{\rm th}e0}}\gg \tau^{-1}_{\rm cool0},
\end{equation}
where the subscript $0$ denotes mean quantities of the hot medium, rather than fluctuations. For short wavelengths of thermally unstable fluctuations (isobaric), the exact growth rate is
\begin{equation}
\label{eq: isob}
\sigma_{\rm ib} =\frac{(2 - \Lambda_{\rm T}) }{\gamma \tau_{\rm cool0}},
\end{equation}
and for long wavelengths (isochoric) the growth rate is
\begin{equation}
\label{eq: isoc}
\sigma_{\rm ic} = - \frac{\Lambda_{\rm T}}{\tau_{\rm cool0}},
\end{equation}
where {\color{black} we use the radiative cooling rate (from eq. \ref{eq:hyd3}) of the hot background medium given by}  $\zeta_0 = \rho^2_0\Lambda(T_0)/{(\mu m_p)}^2$ with $\mu$, the mean molecular mass, and $m_p$, the the proton mass, to evaluate $\Lambda_{\rm T}= \partial \ln \Lambda/\partial \ln T$. Alternatively, we use the timescales $\tau_{\rm ib}$ and $\tau_{\rm ic}$ for isobaric and isochoric growth/decay timescales. The RHS of eq. \ref{eq:clasField} should be the actual growth rate of a given unstable mode, but we ignore that detail in the approximate equations above. The Field length is
\begin{equation}
\label{eq:field}
\lambda_{\rm F} \sim \sqrt{\lambda_{e0}v_{{\rm th}e0} \tau_{\rm cool0}}.
\end{equation}
If $k_{\ell} \beta_e \lambda_e \gg 1$, eq. \ref{eq: anocond} gives the thermal conductivity and the condition on the rates for TI to happen is the following:
\begin{eqnarray}
\label{eq: anoField}
\frac{k_{\ell}v_{{\rm th}e0}}{\beta_{e0}} \ll \tau^{-1}_{\rm cool0},
\end{eqnarray}
or, equivalently, 
\begin{eqnarray}
\frac{\tau_{\rm cool0}}{\tau_{\rm snd0}}\ll \beta_{e0} \sqrt{\frac{m_e}{m_i}}.
\end{eqnarray}
\newline
Therefore, the anomalous TI condition sets a new length scale below which anomalous thermal conduction interferes with TI growth.
\begin{equation}
\label{eq: ano_lmin}
\lambda_{\rm a} \sim \frac{\lambda_{\rm s}}{\beta_{e0}}\sqrt{\frac{m_i}{m_e}}.
\end{equation}
We can write the ratio of the minimum scales for the onset of TI as
\begin{equation}
\label{eq: l_ratio}
\frac{\lambda_{\rm a}}{\lambda_{\rm F}} \sim \frac{668}{\beta_{e0}} {\left(\frac{T}{{{\rm keV}}}\right)}^{-\frac{1}{2}}.
\end{equation}
For a $10~{\rm keV}$ plasma with $\beta_{e0}\sim 1000$, $\lambda_{\rm a}/\lambda_{\rm F}\sim 0.2$. This implies that anomalous conduction could allow scales orders of magnitude below the classical Field length to remain unstable depending on the plasma $\beta_e$. Fig. \ref{fig:fig1} shows the LaTI regime based on the inequality in eq. \ref{eq: anoField}. The yellow region defines the usual TI parameter space. The diagonal line determines the changes in the length scale of a perturbation ($k_0$). The red region shows the parameter space in which the heat-flux-driven whistler instability is active. Within the red area, the purple region defines LaTI. We verify this parameter space by carrying out simulations at different $k_0$ as described in section \ref{sec:sim}.

\begin{figure*}
    \includegraphics[width=18cm]{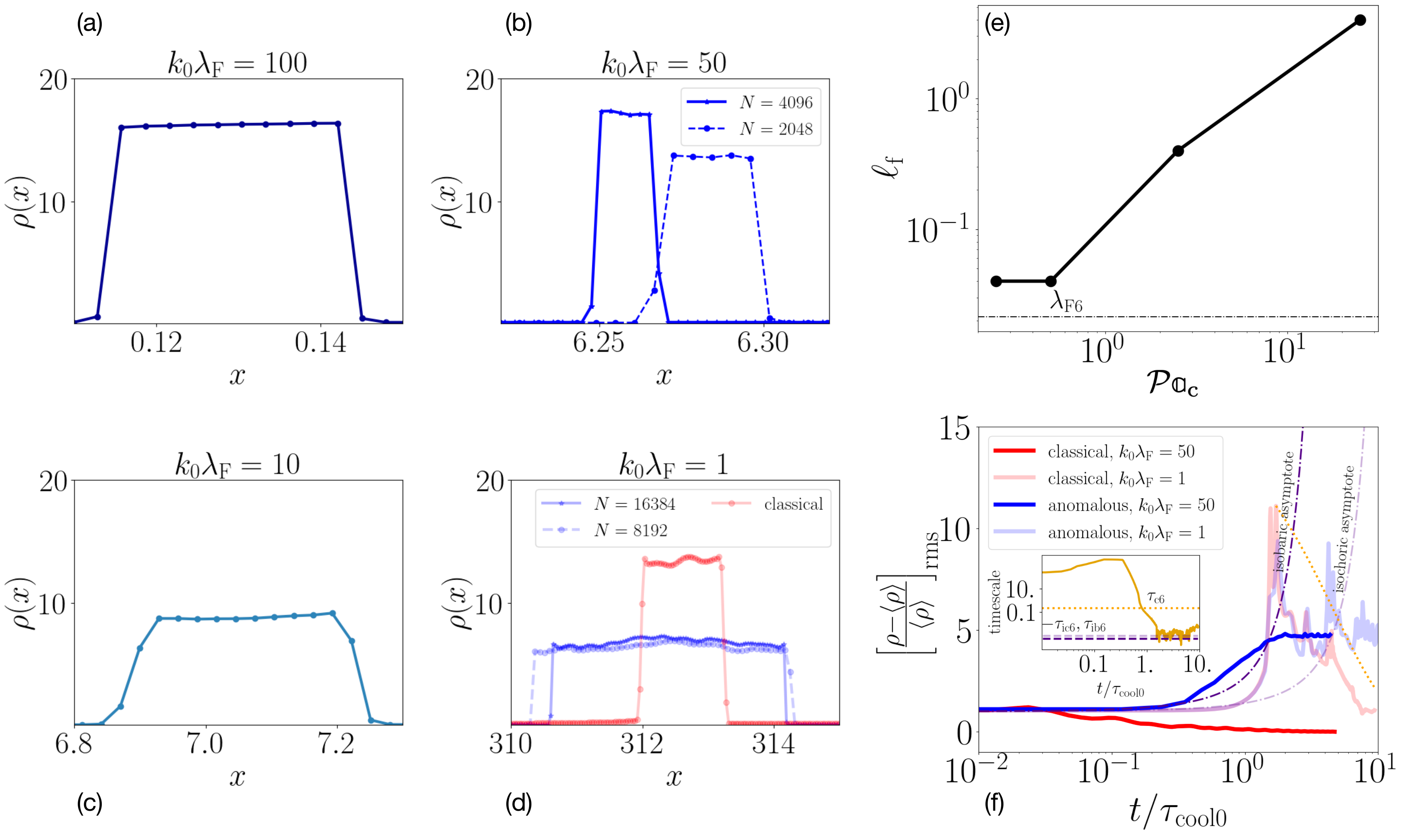}
    \centering
    \caption{The first two columns [(a),(b),(c),(d)] represent four clouds of varying initial size $\ell_{\rm T,\parallel 0} (k_0\lambda_{\rm F})$ after saturation in the simulations with anomalous conductivity, radiative cooling and maintained net thermal balance. {\color{black} The x-axis and y-axis are in code units ({\color{black}${3.086 \times 10^{21}~{\rm cm}}$ and $1.672\times 10^{-24} ~{\rm g~ {cm}^{-3}}$} respectively). The second column [(b), (d)] further demonstrates convergence of the cloud sizes.} The upper panel in the third column (e) shows the final size (${\rm kpc}$) of the clouds as a function of initial ${\mathcal{P}{\mathbb a}}_{\rm c}=\lambda_{\rm s} \ell/\lambda^2_{\rm F}$ and $\lambda_{\rm F6}$ is the predicted Field length at low temperature stable phase ($2\times 10^6~{\rm K}$) in our simulations. The lower panel (f) in the third column shows rms density fluctuations with time (normalised to mean $\tau_{\rm cool}$). In order to understand the decay of the classical large scale cloud (light red), we further show the predicted classical decay timescale {\color{black}($\tau_{\rm c6}$, orange), isobaric TI timescale ($\tau_{\rm ib6}$, dark purple dashed) and isochoric TI timescale ($\tau_{\rm ic6}$, light purple dashed)} at $2\times 10^6$ K across all mean densities (at each time) in the inset. {\color{black} The orange line essentially tracks the decrease in the characteristic cloud size in the box.} }
    \label{fig:fig3}
\end{figure*}

\subsection{What are the saturation mechanisms?}
If we consider energy conservation for the cooling plasma at early times with linearized perturbations ($\Lambda_{\rm T}=0.5$), the terms related to advection, TI, and conduction in the following dominate the energy budget:
\begin{eqnarray}
\nonumber
\sigma_{\rm adv}\sim\frac{\gamma c_{\rm s0}}{\ell}, \sigma_{\rm TI}\sim\frac{3}{2}\tau_{\rm cool0}^{-1}, \sigma_{\rm (c,a)}\sim\frac{\kappa_{\rm (c,a), \parallel} T_0(\gamma-1)}{\ell^2 p_0},
\end{eqnarray}
where $\ell$ is the perturbation length scale and $\kappa_{\parallel (c,a)}$ indicates classical or anomalous conduction. Advection and conduction are the two possible saturation mechanisms competing with the growth term above. We define a dimensionless number, analogous to the thermal {\color{black} Peclet} number, that indicates the likely saturation mechanism for a given initial scale $\ell$,
\begin{equation}
\label{eq: dimlno}
\mathcal{P}{\mathbb a} = \frac{\sigma_{\rm adv}}{\sigma_{\rm (c,a)}} = \frac{p_0 c_{\rm s} \ell}{\kappa_{\rm (c,a), \parallel}T_0(\gamma -1)}. 
\end{equation}
This dimensionless number is ${\mathcal{P}{\mathbb a}}_{\rm c}=\lambda_{\rm s} \ell/\lambda^2_{\rm F}$ and ${\mathcal{P}{\mathbb a}}_{\rm a}=\beta_e \sqrt{m_e/m_i}$ for classical and anomalous conduction, respectively. If $\mathcal{P}{\mathbb a}\ll1$, conduction dominates. At steady state, the perturbations reach high density, low temperature, and zero velocity if they are divergence-free. At the peak of the perturbations, $\partial T/\partial x=0$ and $\partial^2T/\partial x^2\neq0$, so classical conduction may take over and the final size is then set by approximately the classical Field length at the final temperature and density. By contrast, $\mathcal{P}{\mathbb a}\gg1$ implies that advection drives the dynamics, and the final scale is proportional to the initial temperature-perturbation scale before the onset of TI. Classical conduction may still act on the clump edges in this case, but may not set the final length scale.
\section{1D Hydrodynamic simulations}
\label{sec:sim}
We use the {\tt PLUTO} code (\citealt{2007ApJS_mignone}) to solve the hydrodynamic eqs \ref{eq:hyd1}-\ref{eq:hyd3} with parameter $\beta_e$, effectively modeling a cylindrical hot plasma along the magnetic field. We use a super-timestepping scheme for thermal conduction and fix the parabolic Courant number by comparing 1D tests of temperature-fluctuation decay with analytical expectations. We use periodic boundary conditions and a uniform grid size $N$ with the box size $L_0\approx 2.4\ell_{\rm T,\parallel0}$. For the initial condition, we take a simple spatial temperature profile in pressure equilibrium (low density corresponds to high temperature):
\begin{equation}
\label{eq:temp_ini}
T(x) = \bar{T}\left(1- A \sin (k_0 x)\right) . 
\end{equation}
We use $\varepsilon=k_0\lambda_{e0}\beta_{e0}$ as a parameter, and all basic simulation parameters are listed in Table \ref{tab:thetab}. All definitions used in the simulations are identical to those in section \ref{sec: prelim}.
\begin{table}
\label{tab:thetab}
\begin{center}
\begin{tabular}{| c c c c c|}
\hline
$\bar{T},\bar{n}$& $\varepsilon=k_0\lambda_{e0}\beta_{e0}$ & $A, \ell_{{\rm T},\parallel0}$ & C/H& $N$\\
\hline\hline
$1,2.2$&$36.4$&$0.2, 25$& n&$2048$\\
\hline
$1,2.2$&-&$0.2, 25$& n&$2048$\\
\hline
$1,2.2$ &$364$ &$0.2, 25$&n&$2048$\\
\hline
$8,0.1$ &$256$ &$0.2, 25$&n&$2048$\\
\hline
$8,0.9$&-&$0.9, 5$&y&$4096$\\
\hline
$8,0.9$&-&$0.9, 250$&y&$16384$\\
\hline
$8,0.9$& $284$ &$0.9, 2.5$&y&$4096$\\
\hline
$8,0.9$&$142$ &$0.9, 5$&y&$2048$\\
\hline
$8,0.9$&$142$&$0.9, 5$&y&$4096$\\
\hline
$8,0.9$&$28$ &$0.9, 25$&y&$4096$\\
\hline
$8,0.9$&$2.8$ &$0.9, 250$&y&$16384$\\
\hline
$8,0.9$&$2.8$ &$0.9, 250$&y&$8192$\\
\hline
\end{tabular}
\end{center}
\caption{The table with simulation parameters. Temperature is in the unit of $10^7~{\rm K}$, number density is in the unit of $0.1~{\rm cm^{-3}}$, and $\ell_{\rm T,\parallel0}$ is in kpc. All the runs with cooling (C/H) have a $\lambda_{\rm F}\approx 40~{\rm kpc}$ and the cooling time $\tau_{\rm cool0}\approx0.12~{\rm Gyr}$ (although for large $A$, the mean density is not simply $p_0/k_{\rm B}\bar{T}$ using eq. \ref{eq:temp_ini}, as has been used for the $\tau_{\rm cool0}$ estimate). Simulations without $\varepsilon$ specified have classical conduction only. The lowest temperature in simulations with C/H is $2\times 10^6~{\rm K}$ as otherwise it is computationally more expensive. We resolve the classical Field length at $2\times 10^6~{\rm K}$ only for the cases with $\ell_{\rm T,\parallel0}\in [2.5,5]~{\rm kpc}$ but we find that it is not necessary to resolve it for large condensates.}
\end{table}
\subsection{Tests with classical and anomalous decay}
We carry out several tests to compare how classical and anomalous conduction produce distinct temperature-fluctuation evolutions. For a range of $\varepsilon$, we evaluate the root-mean-squared (rms) temperature fluctuation in the simulation box over time in Fig. \ref{fig:fig2} (left panel). We normalise time by the classical decay timescale, $\tau_{\rm c}$. As expected, the decay is slower by a factor proportional to $\varepsilon$. The right panel compares the classical and anomalous simulations for a given $\varepsilon$. The fluctuations decay slowly as predicted, but each coherent temperature structure also shows unique features. The fluctuation peak is flattened relative to classical conduction as $\partial T/\partial x$ approaches zero and classical conduction is activated. However, near the bottom of a temperature peak, $\partial T/\partial x$ is larger, leading to the maintenance of this approximate step function.
\subsection{Tests with radiative cooling}

We carry out several simulations with cooling/heating as described in section \ref{sec: prelim}. Thermal instability causes runaway cooling for dense clumps, and the resulting scale separation in the non-linear stage is especially challenging to capture when thermal conduction is included. Therefore, we impose a temperature ceiling ($2\times 10^8~{\rm K}$) and floor ($2\times 10^6~{\rm K}$) in our simulations. While this approach artificially flattens temperature peaks and troughs, it also reflects the behavior of a stable-phase plasma in which the TI growth rate is small. In fact, at $2\times 10^6~{\rm K}$, TI may not exist, depending on the plasma density (as characterized for a given density in Fig. 6 by \citealt{Das2021ShatterInstability}).

We design four simulations with modified conduction over a range of $k_0$, as given in Table \ref{tab:thetab} and shown in Fig. \ref{fig:fig1}, spanning the LaTI regime. We also carry out two equivalent simulations with $\ell_{\rm T,\parallel0}\in [5,250]~{\rm kpc}$ using only classical conduction for comparison. The first two columns in Fig. \ref{fig:fig3}, [(a),(b),(c),[d)], show the four condensates after saturation in the LaTI regime {\color{black}(and in two cases, (b) and (d), we show different resolutions to demonstrate convergence)}. For the two smaller perturbation sizes of the four, the initial ${\mathcal{P}{\mathbb a}}_{\rm c}$, {\color{black} the classical thermal {\color{black} Peclet} number}, is less than unity; in other words, advection may not strongly oppose cooling, but thermal conduction will. The final size $\ell_{\rm f}$ of the condensate is ${\sim}\lambda_{\rm F6}$, the classical Field length at $2\times 10^6~{\rm K}$ (upper panel, (e), of the last column). Large-scale perturbations scale down such that $\ell_{\rm f}/\ell_{\rm T,\parallel0}$ is constant in both cases. Classical conduction does not contribute significantly to the final size {\color{black} of these large clouds. A key conclusion is that the cloud formation is triggered in the LaTI regime, but the saturation progresses as in the usual TI and the role of anomalous conduction is only to insulate the cold phase from the hot phase}.

The lower panel, (f), of the last column shows the rms density fluctuations over time (normalised to initial $\tau_{\rm cool0}$). The dark blue line, which shows LaTI for $k_0\lambda_{\rm F}=50$, clearly grows to amplitudes as large as the nearly isochoric cases (light blue and red). The onset closely matches the isobaric TI, as expected (also see Fig. \ref{fig:fig1}). For the classical and anomalous cases with $k_0\lambda_{\rm F}=1$, a distinct feature is that classical conduction causes the amplitude to decay after a few cooling times. This decay rate is approximately equal to the inverse of $0.2~{\rm Gyr}$ (orange dotted line). To verify this in simulation, we plot in the inset the predicted timescales $\tau_{\rm c6}, \tau_{\rm ib6}, -\tau_{\rm ic6}$ corresponding to classical decay (orange), isobaric growth (dark purple dashed), and isochoric growth (decay; light purple dashed) at $2\times 10^6~{\rm K}$ across the mean density in the box at all times. {\color{black} Note that $\tau_{\rm c6}$ essentially tracks a change in the characteristic size of the clouds in the box.} These predicted rates are only relevant beyond $t\gtrsim \tau_{\rm cool0}$, when the classical decay timescale drops rapidly as the condensate shrinks, until it equals the isobaric timescale. Therefore, even in the traditional TI regime, anomalous conduction insulates saturated large-scale condensates against mixing with the surrounding medium.
\section{Implications for cluster outskirts}
\label{sec:cluster_disc}
\subsection{LaTI in an isothermal ICM}
We consider an isothermal model of the galaxy cluster atmosphere to assess at which radii LaTI can operate. We denote all variables at the cluster's virial radius with the subscript `${\rm vir}$'. The virial temperature and virial radius ($R_{\rm vir}$) scale with the total mass of the cluster's dark matter halo ($M_{\rm vir}$). 
\begin{eqnarray}
\label{eq:Tvir}
    \left(\frac{T_{\rm vir}}{\rm keV}\right) &=& \frac{2\mu m_p G}{3k_{\rm B}{\rm keV}}{\left[\frac{4}{3}\pi (200\rho_{\rm crit}) \right]}R^2_{\rm vir} \\
    \nonumber
    &=& 2.3{\left(\frac{R_{\rm vir}}{\rm Mpc}\right)}^2,
\end{eqnarray}
where $G$ is the gravitational constant, and $\rho_{\rm crit}=10^{-29}~{\rm g~cm^{-3}}$ is the critical density of the Universe at the current time.
For $T_{\rm vir}=4~{\rm keV}$, we find $R_{\rm vir}=1.3~{\rm Mpc}$. The temperature is constant throughout the ICM, and the density falls off as the square of radius. Using flux freezing for the magnetic field, it follows that
\begin{equation}
\label{eq:virial}
\rho_0 (R) = \frac{\rho_{\rm vir} R^2_{\rm vir}}{R^2}, \, T_0(R) = T_{\rm vir}, \, B_0(R) \propto \rho_0^{{2}/{3}}(R).
\end{equation}
\newline
Using the above, the classical and anomalous TI growth conditions (inequalities \ref{eq:clasField} and \ref{eq: anoField}) become, respectively, 
\begin{eqnarray}
\label{eq:vir_cond1}
\frac{R^2}{R^2_{\rm vir}} &\ll& \frac{1}{k_{\ell}\lambda_{\rm F, vir}},\\
\label{eq:vir_cond2}
\frac{R^2}{R^2_{\rm vir}} &\ll& 3.2\times10^{-5}\frac{{\beta_{e,{\rm vir}}}^{{3}/{2}}}{{\left(k_{\ell}\lambda_{\rm F, vir}\right)}^{{3}/{2}}}{\left(\frac{T_{\rm vir}}{\rm keV}\right)}^{{3}/{4}}.
\end{eqnarray}
\newline
\newline
The second inequality applies only if conduction is anomalously suppressed, which requires
\begin{equation}
\label{eq:anom_vir}
\frac{R^2}{R^2_{\rm vir}} \gg \frac{145}{{\left(k_{\ell}\lambda_{\rm F,vir} \right)}^{{3}/{4}}}{\beta_{e,{\rm vir}}^{-3/4}\left(\frac{T_{\rm vir}}{\rm keV}\right)}^{-{3}/{8}} . 
\end{equation}
Additionally, these conditions apply only when a TI mode fits within a given radius, so $k_{\ell} R_{\rm max}\sim k_{\ell}\lambda_{\rm F}$ and $k_{\ell} R_{\rm max}\sim k_{\ell}\lambda_{\rm m}$ set the two feasibility conditions for the two conductivities. This gives the following scalings for the maximum radius; in the anomalous case, it depends on the virial temperature and plasma $\beta_{e,{\rm vir}}$:
\begin{eqnarray}
\label{eq:Rm_cond1}
\frac{R_{\rm max, c}}{R_{\rm vir}} &=& \frac{R_{\rm vir}}{\lambda_{\rm F, vir}}\sim 0.3{\left(\frac{T_{\rm vir}}{\rm keV}\right)}^{-1}, \\
\label{eq:Rm_cond2}
\frac{R_{\rm max, a}}{R_{\rm vir}} &=& 10^{-9}{{\beta_{e,{\rm vir}}}^{3}}{\left(\frac{R_{\rm vir}}{\lambda_{\rm F, vir}}\right)}^{3}{\left(\frac{T_{\rm vir}}{\rm keV}\right)}^{{3/}{2}}.
\end{eqnarray}
\newline
Fig. \ref{fig:fig4} shows these conditions in parameter space for a galaxy cluster, with fractional radial extent on the $y$-axis and size relative to the Field length at the virial radius on the $x$-axis. The orange region between the solid thin and the dashed orange lines, respectively, is the classical TI growth regime constrained by mode-size limits at any radius. This lies within a relatively small inner fraction of the ICM ($R_{\rm max,c}/R_{\rm vir} \approx 0.07$). The solid indigo line marks the LaTI condition, and the red line marks the anomalous-effect condition. The purple shaded region is the LaTI regime, extending to $\gtrsim50\%$ of the cluster ($R_{\rm max,a}/R_{\rm vir} \approx 0.5$) and overlapping with the classical regime (excluding about $5\%$ from the center). From eqs. \ref{eq:Rm_cond1}-\ref{eq:Rm_cond2}, we note that the constraint on fitting an unstable mode within a radius depends on the hot plasma's temperature and may enhance or reduce the LaTI radial extent. In addition, the LaTI regime is expanded for larger plasma $\beta_e$ due to suppression of heat flux by a factor ${\sim}\beta^{-1}_e$.
This is true for all $\beta_e \lesssim \ell_{\rm T,\parallel}/r_e$ (\citealt{Davies26}), which is an exceptionally weak constraint for the smallest magnetic field in the ICM ($\ell_{\rm T,\parallel}/r_e \sim 10^{12}{-}10^{14}$) to trigger heat-flux suppression.
\subsection{How does LaTI compete with turbulent stirring?}
\begin{figure}
\smallskip
    \includegraphics[width=8cm]{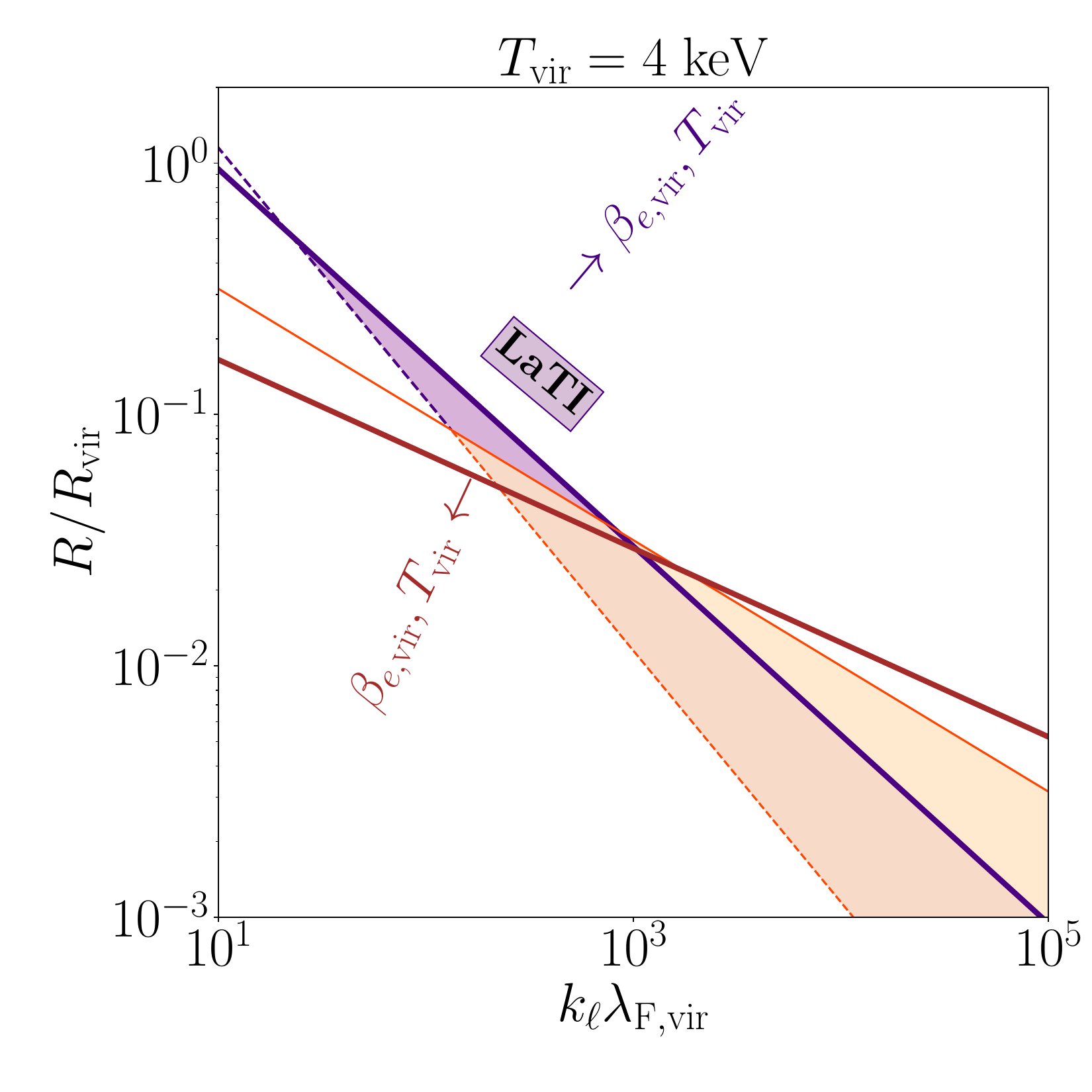}
    \centering
    \caption{The parametric space of an isothermal intracluster medium in which the classical conduction limited TI (light yellow) and the LaTI regime (light purple below the solid indigo line) are shown. The classical conduction limited TI must occur between orange solid and orange dashed lines where the latter marks the size limit of TI modes for a given radius ($k_\ell R\sim 1$). The solid red line marks the anomalous conduction regime $k_{\ell}\beta_e \lambda_e\sim 1$. The purple and red lines move up and down respectively with higher $T_{\rm vir}$ and $\beta_e$.}
    \label{fig:fig4}
\end{figure}

A key question in the context of LaTI in the ICM is whether LaTI can produce abundant small-scale temperature fluctuations instead of turbulent stirring. This addresses the problem of persistent, steady-state temperature fluctuations in the ICM. We consider an outer scale $\ell_0$ at which energy is injected, and a second scale $\ell_{\rm C}$ at which the energy-injection rate equals the cooling rate. Using Kolmogorov scaling for turbulent velocities, we equate the eddy turnover timescale to the cooling timescale to determine $\ell_{\rm C}$: 
\begin{eqnarray}
\frac{\ell_{\rm C}}{u_{\rm C}} &\sim& \frac{\ell_0}{u_0} {\left( \frac{\ell_{\rm C}}{\ell_0}\right)}^{{2}/{3}} \sim \tau_{\rm cool},
\end{eqnarray}
which gives
\begin{eqnarray}
\label{eq:turb}
\frac{\ell_{\rm C}}{\ell_0} &\sim& {\left( \frac{\tau_{\rm cool}}{\tau_{{\rm snd},\ell_0}} \mathcal{M}\right)}^{{3}/{2}} \propto R^{3/2}. 
\end{eqnarray}
We note that the classical Field length $\propto R^2$ and the constraint on the size of unstable modes $k^{-1}_{\ell}\propto R$, therefore $\ell_{\rm C}$ surpass the constraint on size at a smaller cluster radius than where it reaches the Field length. Hence $\ell_{\rm C}\sim \ell_0\sim R_{\rm turb}$ sets the radius beyond which turbulence is dominant over condensation since the size constraint resists LaTI. For a $4~{\rm keV}$ plasma as in Fig. \ref{fig:fig4}, we find $R_{\rm turb}\lesssim R_{\rm max,c}$. Although the onset of LaTI could be affected by turbulence beyond this radius, multi-dimensional numerical simulations show that turbulence and TI can enhance each other non-linearly (\citealt{2016MNRAS_choudhury, Mohapatra2019TurbulenceThermodynamics}). 
\section{Conclusions}
\label{sec:disc}
We propose a new mechanism for producing temperature fluctuations outside the core of the intracluster medium. X-ray surface-brightness and SZ pressure maps suggest that clusters are permeated by persistent density, pressure, and temperature fluctuations. Thermal instability, or Field instability, can operate only within $\sim 7{-}8\%$ of the cluster core to form low-temperature condensates, since thermal conduction damps fluctuations beyond that extent. However, much of the region outside the core has a collisional mean free path large relative to the temperature-gradient scale, and a weak magnetic field can easily trigger microinstabilities, specifically heat-flux-driven whistlers, that suppress heat flux. This opens a new parametric regime, which we call `Latent thermal instability' (LaTI), in which unstable modes can grow across nearly $20-70\%$ of the cluster, depending on the hot medium's temperature, and provide a viable mechanism for steady-state isobaric and isochoric fluctuations throughout this radial range. We use high-resolution 1D hydrodynamic cloud simulations to test our analytical prediction. 

{\color{black} A limitation of this current research is that, due to the simplified 1D geometry we adopt, we do not capture magnetic-field amplification inside a condensate due to adiabatic compression. This should not change the destabilization criteria for LaTI, but the lower $\beta_e$ in the condensate may interfere with its saturation. However, the most relevant $\beta_e$ for the anomalous heat flux should be at the interface of the condensate and the hot medium rather than within the condensate, because it is the temperature gradient at this interface that provides the free energy to destabilize whistlers. Therefore, magnetization in this mixed medium should play a key role in saturation. We further note that compression itself can destabilize other kinetic instabilities that suppress heat-flux by a factor dependent on the compression rate and the local $\beta_e$, thus sustaining the fluctuations, e.g., ~\citealt{Komarov_2016}. Testing this proposed mechanism in multi-dimensional magnetohydrodynamic models of galaxy clusters is an essential future research direction.}

\begin{acknowledgments}
PPC acknowledges a computational resource award from the DiRAC High Performance Computing facility. This work used the DiRAC Data Intensive service (CSD3) at the University of Cambridge, managed by the University of Cambridge University Information Services on behalf of the STFC DiRAC HPC Facility (\href{www.dirac.ac.uk}{www.dirac.ac.uk}). The DiRAC component of CSD3 at Cambridge was funded by BEIS, UKRI and STFC capital funding and STFC operations grants. DiRAC is part of the UKRI Digital Research Infrastructure. PPC and AFAB have also been primarily supported by UKRI (grant number MR/W006723/1) during the course of this research.
\end{acknowledgments}

\begin{contribution}
PPC conceptualised the problem, performed analytical calculations, simulations, data analysis, and wrote the manuscript. AFAB actively participated in discussions on the results, and final stage of the manuscript. 
\end{contribution}

%






\bibliography{sample7}{}

@ARTICLE{2018JPlPh_komarov,
       author = {{Komarov}, S. and {Schekochihin}, A.~A. and {Churazov}, E. and {Spitkovsky}, A.},
        title = "{Self-inhibiting thermal conduction in a high- , whistler-unstable plasma}",
      journal = {Journal of Plasma Physics},
         year = 2018,
        month = jun,
       volume = {84},
       number = {3},
          eid = {905840305},
        pages = {905840305},
          doi = {10.1017/S0022377818000399},
archivePrefix = {arXiv},
       eprint = {1711.11462},
 primaryClass = {astro-ph.HE},
       adsurl = {https://ui.adsabs.harvard.edu/abs/2018JPlPh..84c9005K}
}

@ARTICLE{2024A&A_perrone,
       author = {{Perrone}, Lorenzo Maria and {Berlok}, Thomas and {Pfrommer}, Christoph},
        title = "{Thermal conductivity with bells and whistlers: Suppression of the magnetothermal instability in galaxy clusters}",
      journal = {\aap},
         year = 2024,
        month = oct,
       volume = {690},
          eid = {A292},
        pages = {A292},
          doi = {10.1051/0004-6361/202449555},
archivePrefix = {arXiv},
       eprint = {2402.06718},
 primaryClass = {astro-ph.CO},
       adsurl = {https://ui.adsabs.harvard.edu/abs/2024A&A...690A.292P}
}

@ARTICLE{2022A&A_beckmann,
       author = {{Beckmann}, Ricarda S. and {Dubois}, Yohan and {Pellissier}, Alisson and {Polles}, Fiorella L. and {Olivares}, Valeria},
        title = "{AGN jets do not prevent the suppression of conduction by the heat buoyancy instability in simulated galaxy clusters}",
      journal = {\aap},
         year = 2022,
        month = oct,
       volume = {666},
          eid = {A71},
        pages = {A71},
          doi = {10.1051/0004-6361/202243873},
archivePrefix = {arXiv},
       eprint = {2204.12514},
 primaryClass = {astro-ph.GA},
       adsurl = {https://ui.adsabs.harvard.edu/abs/2022A&A...666A..71B}
}

@ARTICLE{1965ApJ_field,
       author = {{Field}, George B.},
        title = "{Thermal Instability.}",
      journal = {\apj},
         year = 1965,
        month = aug,
       volume = {142},
        pages = {531},
          doi = {10.1086/148317},
       adsurl = {https://ui.adsabs.harvard.edu/abs/1965ApJ...142..531F}
}

@article{Das2021ShatterInstability,
    title = {{Shatter or not: Role of temperature and metallicity in the evolution of thermal instability}},
    year = {2021},
    journal = {Monthly Notices of the Royal Astronomical Society},
    author = {Das, Hitesh Kishore and Choudhury, Prakriti Pal and Sharma, Prateek},
    number = {4},
    pages = {4935--4952},
    volume = {502},
    publisher = {Oxford University Press},
    doi = {10.1093/mnras/stab382},
    issn = {13652966},
    arxivId = {2009.11317}
}

@article{Sharma2010ThermalClusters,
    title = {{Thermal instability with anisotropic thermal conduction and adiabatic cosmic rays: Implications for cold filaments in galaxy clusters}},
    year = {2010},
    journal = {Astrophysical Journal},
    author = {Sharma, Prateek and Parrish, Ian J. and Quataert, Eliot},
    number = {1},
    pages = {652--665},
    volume = {720},
    doi = {10.1088/0004-637X/720/1/652},
    issn = {15384357},
    arxivId = {1003.5546}
}

@ARTICLE{1953ApJ_parker,
       author = {{Parker}, Eugene N.},
        title = "{Instability of Thermal Fields.}",
      journal = {\apj},
         year = 1953,
        month = may,
       volume = {117},
        pages = {431},
          doi = {10.1086/145707},
       adsurl = {https://ui.adsabs.harvard.edu/abs/1953ApJ...117..431P}
}

@ARTICLE{2016MNRAS_choudhury,
       author = {{Choudhury}, Prakriti Pal and {Sharma}, Prateek},
        title = "{Cold gas in cluster cores: global stability analysis and non-linear simulations of thermal instability}",
      journal = {\mnras},
         year = 2016,
        month = apr,
       volume = {457},
       number = {3},
        pages = {2554-2568},
          doi = {10.1093/mnras/stw152},
archivePrefix = {arXiv},
       eprint = {1512.01217},
 primaryClass = {astro-ph.GA},
       adsurl = {https://ui.adsabs.harvard.edu/abs/2016MNRAS.457.2554C}
}

@BOOK{1962pfig.book_spitzer,
       author = {{Spitzer}, L.},
        title = "{Physics of Fully Ionized Gases}",
         year = 1962,
       adsurl = {https://ui.adsabs.harvard.edu/abs/1962pfig.book.....S}
}

@ARTICLE{1965RvPP_braginskii,
       author = {{Braginskii}, S.~I.},
        title = "{Transport Processes in a Plasma}",
      journal = {Reviews of Plasma Physics},
         year = 1965,
        month = jan,
       volume = {1},
        pages = {205},
       adsurl = {https://ui.adsabs.harvard.edu/abs/1965RvPP....1..205B}
}

@ARTICLE{2001ApJ_narayan,
       author = {{Narayan}, Ramesh and {Medvedev}, Mikhail V.},
        title = "{Thermal Conduction in Clusters of Galaxies}",
      journal = {\apjl},
         year = 2001,
        month = dec,
       volume = {562},
       number = {2},
        pages = {L129-L132},
          doi = {10.1086/338325},
archivePrefix = {arXiv},
       eprint = {astro-ph/0110567},
 primaryClass = {astro-ph},
       adsurl = {https://ui.adsabs.harvard.edu/abs/2001ApJ...562L.129N}
}

@ARTICLE{1994ARA&A_fabian,
       author = {{Fabian}, A.~C.},
        title = "{Cooling Flows in Clusters of Galaxies}",
      journal = {\araa},
         year = 1994,
        month = jan,
       volume = {32},
        pages = {277-318},
          doi = {10.1146/annurev.aa.32.090194.001425},
       adsurl = {https://ui.adsabs.harvard.edu/abs/1994ARA&A..32..277F}
}

@ARTICLE{2018MNRAS_ji,
       author = {{Ji}, Suoqing and {Oh}, S. Peng and {McCourt}, Michael},
        title = "{The impact of magnetic fields on thermal instability}",
      journal = {\mnras},
         year = 2018,
        month = may,
       volume = {476},
       number = {1},
        pages = {852-867},
          doi = {10.1093/mnras/sty293},
archivePrefix = {arXiv},
       eprint = {1710.00822},
 primaryClass = {astro-ph.GA},
       adsurl = {https://ui.adsabs.harvard.edu/abs/2018MNRAS.476..852J}
}

@ARTICLE{2019A&A_claes,
       author = {{Claes}, N. and {Keppens}, R.},
        title = "{Thermal stability of magnetohydrodynamic modes in homogeneous plasmas}",
      journal = {\aap},
         year = 2019,
        month = apr,
       volume = {624},
          eid = {A96},
        pages = {A96},
          doi = {10.1051/0004-6361/201834699},
       adsurl = {https://ui.adsabs.harvard.edu/abs/2019A&A...624A..96C}
}

@ARTICLE{2025MNRAS_choudhury,
       author = {{Choudhury}, Prakriti Pal and {Reynolds}, Christopher S.},
        title = "{Cold fronts in galaxy clusters - I. A case for the large-scale global eigenmodes in unmagnetized and weakly magnetized cluster core}",
      journal = {\mnras},
         year = 2025,
        month = mar,
       volume = {537},
       number = {4},
        pages = {3194-3209},
          doi = {10.1093/mnras/staf184},
archivePrefix = {arXiv},
       eprint = {2408.03988},
 primaryClass = {astro-ph.GA},
       adsurl = {https://ui.adsabs.harvard.edu/abs/2025MNRAS.537.3194C}
}

@ARTICLE{2025MNRAS_kaul,
       author = {{Kaul}, Ish and {Tan}, Brent and {Oh}, S. Peng and {Mandelker}, Nir},
        title = "{Tales of tension: magnetized infalling cold clouds and streams in the CGM}",
      journal = {\mnras},
         year = 2025,
        month = jun,
       volume = {539},
       number = {4},
        pages = {3669-3696},
          doi = {10.1093/mnras/staf706},
archivePrefix = {arXiv},
       eprint = {2502.17549},
 primaryClass = {astro-ph.GA},
       adsurl = {https://ui.adsabs.harvard.edu/abs/2025MNRAS.539.3669K}
}

@ARTICLE{2026arXiv_voit,
       author = {{Voit}, G.~M. and {Wibking}, B.~D. and {Yaldiz}, D.},
        title = "{Magnetohydrodynamic Precipitation}",
      journal = {arXiv e-prints},
         year = 2026,
        month = feb,
          eid = {arXiv:2602.15121},
        pages = {arXiv:2602.15121},
          doi = {10.48550/arXiv.2602.15121},
archivePrefix = {arXiv},
       eprint = {2602.15121},
 primaryClass = {astro-ph.GA},
       adsurl = {https://ui.adsabs.harvard.edu/abs/2026arXiv260215121V}
}

@ARTICLE{2003ApJ_zakamska,
       author = {{Zakamska}, Nadia L. and {Narayan}, Ramesh},
        title = "{Models of Galaxy Clusters with Thermal Conduction}",
      journal = {\apj},
         year = 2003,
        month = jan,
       volume = {582},
       number = {1},
        pages = {162-169},
          doi = {10.1086/344641},
archivePrefix = {arXiv},
       eprint = {astro-ph/0207127},
 primaryClass = {astro-ph},
       adsurl = {https://ui.adsabs.harvard.edu/abs/2003ApJ...582..162Z}
}

@ARTICLE{2019MNRAS_choudhury,
       author = {{Choudhury}, Prakriti Pal and {Sharma}, Prateek and {Quataert}, Eliot},
        title = "{Multiphase gas in the circumgalactic medium: relative role of t$_{cool}$/t$_{ff}$ and density fluctuations}",
      journal = {\mnras},
         year = 2019,
        month = sep,
       volume = {488},
       number = {3},
        pages = {3195-3210},
          doi = {10.1093/mnras/stz1857},
archivePrefix = {arXiv},
       eprint = {1901.02903},
 primaryClass = {astro-ph.GA},
       adsurl = {https://ui.adsabs.harvard.edu/abs/2019MNRAS.488.3195C}
}

@ARTICLE{2009ApJ_parrish,
       author = {{Parrish}, Ian J. and {Quataert}, Eliot and {Sharma}, Prateek},
        title = "{Anisotropic Thermal Conduction and the Cooling Flow Problem in Galaxy Clusters}",
      journal = {\apj},
         year = 2009,
        month = sep,
       volume = {703},
       number = {1},
        pages = {96-108},
          doi = {10.1088/0004-637X/703/1/96},
archivePrefix = {arXiv},
       eprint = {0905.4500},
 primaryClass = {astro-ph.CO},
       adsurl = {https://ui.adsabs.harvard.edu/abs/2009ApJ...703...96P}
}

@ARTICLE{1999A&A_kull,
       author = {{Kull}, A. and {B{\"o}hringer}, H.},
        title = "{Detection of filamentary X-ray structure in the core of the Shapley supercluster}",
      journal = {\aap},
         year = 1999,
        month = jan,
       volume = {341},
        pages = {23-28},
          doi = {10.48550/arXiv.astro-ph/9812319},
archivePrefix = {arXiv},
       eprint = {astro-ph/9812319},
 primaryClass = {astro-ph},
       adsurl = {https://ui.adsabs.harvard.edu/abs/1999A&A...341...23K}
}

@ARTICLE{1980MNRAS_cowie,
       author = {{Cowie}, L.~L. and {Fabian}, A.~C. and {Nulsen}, P.~E.~J.},
        title = "{NGC 1275 and the Perseus cluster - The formation of optical filaments in cooling gas in X-ray clusters}",
      journal = {\mnras},
         year = 1980,
        month = may,
       volume = {191},
        pages = {399-410},
          doi = {10.1093/mnras/191.2.399},
       adsurl = {https://ui.adsabs.harvard.edu/abs/1980MNRAS.191..399C}
}

@ARTICLE{1995A&A_briel,
       author = {{Briel}, U.~G. and {Henry}, J.~P.},
        title = "{Search for X-ray filaments between galaxy clusters.}",
      journal = {\aap},
         year = 1995,
        month = oct,
       volume = {302},
        pages = {L9},
       adsurl = {https://ui.adsabs.harvard.edu/abs/1995A&A...302L...9B}
}

@ARTICLE{2004ApJ_sparks,
       author = {{Sparks}, William B. and {Donahue}, Megan and {Jord{\'a}n}, Andr{\'e}s and {Ferrarese}, Laura and {C{\^o}t{\'e}}, Patrick},
        title = "{X-Ray and Optical Filaments in M87}",
      journal = {\apj},
         year = 2004,
        month = may,
       volume = {607},
       number = {1},
        pages = {294-301},
          doi = {10.1086/383189},
archivePrefix = {arXiv},
       eprint = {astro-ph/0402204},
 primaryClass = {astro-ph},
       adsurl = {https://ui.adsabs.harvard.edu/abs/2004ApJ...607..294S}
}

@ARTICLE{2006ApJ_ascasibar,
       author = {{Ascasibar}, Yago and {Markevitch}, Maxim},
        title = "{The Origin of Cold Fronts in the Cores of Relaxed Galaxy Clusters}",
      journal = {\apj},
         year = 2006,
        month = oct,
       volume = {650},
       number = {1},
        pages = {102-127},
          doi = {10.1086/506508},
archivePrefix = {arXiv},
       eprint = {astro-ph/0603246},
 primaryClass = {astro-ph},
       adsurl = {https://ui.adsabs.harvard.edu/abs/2006ApJ...650..102A}
}

@ARTICLE{2004MNRAS_nipoti,
       author = {{Nipoti}, Carlo and {Binney}, James},
        title = "{Cold filaments in galaxy clusters: effects of heat conduction}",
      journal = {\mnras},
         year = 2004,
        month = apr,
       volume = {349},
       number = {4},
        pages = {1509-1515},
          doi = {10.1111/j.1365-2966.2004.07628.x},
archivePrefix = {arXiv},
       eprint = {astro-ph/0401106},
 primaryClass = {astro-ph},
       adsurl = {https://ui.adsabs.harvard.edu/abs/2004MNRAS.349.1509N}
}

@techreport{Cowie1977THERATES,
    title = {{THE EVAPORATION OF SPHERICAL CLOUDS IN A HOT GAS. I. CLASSICAL AND SATURATED MASS LOSS RATES}},
    year = {1977},
    author = {Cowie, Lennox L and Mckee, Christopher F},
    pages = {135--146},
    volume = {211}
}

@article{levinson1992inhibition,
  title={Inhibition of electron thermal conduction by electromagnetic instabilities},
  author={Levinson, Amir and Eichler, David},
  journal={Astrophysical Journal, Part 1 (ISSN 0004-637X), vol. 387, March 1, 1992, p. 212-218. Research supported by NASA and USIBSF.},
  volume={387},
  pages={212--218},
  year={1992}
}

@article{Pistinner1998,
    title = {{Self-inhibiting heat flux}},
    year = {1998},
    journal = {Monthly Notices of the Royal Astronomical Society},
    author = {Pistinner, S. L. and Eichler, D.},
    number = {1},
    month = {11},
    pages = {49--58},
    volume = {301},
    publisher = {Blackwell Publishing Ltd},
    doi = {10.1046/j.1365-8711.1998.01770.x},
    issn = {00358711},
    arxivId = {astro-ph/9807025}
}

@article{Roberg-Clark2018SuppressionGradient,
    title = {{Suppression of Electron Thermal Conduction by Whistler Turbulence in a Sustained Thermal Gradient}},
    year = {2018},
    journal = {Physical Review Letters},
    author = {Roberg-Clark, G. T. and Drake, J. F. and Reynolds, C. S. and Swisdak, M.},
    number = {3},
    pages = {35101},
    volume = {120},
    publisher = {American Physical Society},
    url = {https://doi.org/10.1103/PhysRevLett.120.035101},
    doi = {10.1103/PhysRevLett.120.035101},
    issn = {10797114},
    pmid = {29400540},
    arxivId = {1709.00057}
}

@article{Yerger2024CollisionlessTurbulence,
    title = {{Collisionless conduction in a high-beta plasma: a collision operator for whistler turbulence}},
    year = {2024},
    author = {Yerger, Evan L. and Kunz, Matthew W. and Bott, Archie F. A. and Spitkovsky, Anatoly},
    month = {5},
    url = {http://arxiv.org/abs/2405.06481},
    arxivId = {2405.06481}
}

@ARTICLE{2025arXiv_choudhury,
       author = {{Choudhury}, Prakriti Pal and {Bott}, Archie F.~A.},
        title = "{Modeling transport in weakly collisional plasmas using thermodynamic forcing}",
      journal = {arXiv e-prints},
         year = 2025,
        month = apr,
          eid = {arXiv:2504.14000},
        pages = {arXiv:2504.14000},
          doi = {10.48550/arXiv.2504.14000},
archivePrefix = {arXiv},
       eprint = {2504.14000},
 primaryClass = {physics.plasm-ph},
       adsurl = {https://ui.adsabs.harvard.edu/abs/2025arXiv250414000C}
}

@article{walker2019physics,
  title={The physics of galaxy cluster outskirts},
  author={Walker, Stephen and Simionescu, Aurora and Nagai, Daisuke and Okabe, Nobuhiro and Eckert, Dominique and Mroczkowski, Tony and Akamatsu, Hiroki and Ettori, Stefano and Ghirardini, Vittorio},
  journal={Space Science Reviews},
  volume={215},
  number={1},
  pages={7},
  year={2019},
  publisher={Springer}
}

@ARTICLE{2016MNRAS_khatri,
       author = {{Khatri}, Rishi and {Gaspari}, Massimo},
        title = "{Thermal SZ fluctuations in the ICM: probing turbulence and thermodynamics in Coma cluster with Planck}",
      journal = {\mnras},
         year = 2016,
        month = nov,
       volume = {463},
       number = {1},
        pages = {655-669},
          doi = {10.1093/mnras/stw2027},
archivePrefix = {arXiv},
       eprint = {1604.03106},
 primaryClass = {astro-ph.CO},
       adsurl = {https://ui.adsabs.harvard.edu/abs/2016MNRAS.463..655K}
}

@ARTICLE{2020MNRAS_mirakhor,
       author = {{Mirakhor}, M.~S. and {Walker}, S.~A.},
        title = "{A complete view of the outskirts of the Coma cluster}",
      journal = {\mnras},
         year = 2020,
        month = sep,
       volume = {497},
       number = {3},
        pages = {3204-3220},
          doi = {10.1093/mnras/staa2203},
archivePrefix = {arXiv},
       eprint = {2007.12194},
 primaryClass = {astro-ph.CO},
       adsurl = {https://ui.adsabs.harvard.edu/abs/2020MNRAS.497.3204M}
}

@ARTICLE{2012MNRAS_churazov,
       author = {{Churazov}, E. and {Vikhlinin}, A. and {Zhuravleva}, I. and {Schekochihin}, A. and {Parrish}, I. and {Sunyaev}, R. and {Forman}, W. and {B{\"o}hringer}, H. and {Randall}, S.},
        title = "{X-ray surface brightness and gas density fluctuations in the Coma cluster}",
      journal = {\mnras},
         year = 2012,
        month = apr,
       volume = {421},
       number = {2},
        pages = {1123-1135},
          doi = {10.1111/j.1365-2966.2011.20372.x},
archivePrefix = {arXiv},
       eprint = {1110.5875},
 primaryClass = {astro-ph.CO},
       adsurl = {https://ui.adsabs.harvard.edu/abs/2012MNRAS.421.1123C}
}

@ARTICLE{2004A&A_schuecker,
       author = {{Schuecker}, P. and {Finoguenov}, A. and {Miniati}, F. and {B{\"o}hringer}, H. and {Briel}, U.~G.},
        title = "{Probing turbulence in the Coma galaxy cluster}",
      journal = {\aap},
         year = 2004,
        month = nov,
       volume = {426},
        pages = {387-397},
          doi = {10.1051/0004-6361:20041039},
archivePrefix = {arXiv},
       eprint = {astro-ph/0404132},
 primaryClass = {astro-ph},
       adsurl = {https://ui.adsabs.harvard.edu/abs/2004A&A...426..387S}
}

@ARTICLE{2003ApJ_markevitch,
       author = {{Markevitch}, M. and {Mazzotta}, P. and {Vikhlinin}, A. and {Burke}, D. and {Butt}, Y. and {David}, L. and {Donnelly}, H. and {Forman}, W.~R. and {Harris}, D. and {Kim}, D.-W. and {Virani}, S. and {Vrtilek}, J.},
        title = "{Chandra Temperature Map of A754 and Constraints on Thermal Conduction}",
      journal = {\apjl},
         year = 2003,
        month = mar,
       volume = {586},
       number = {1},
        pages = {L19-L23},
          doi = {10.1086/374656},
archivePrefix = {arXiv},
       eprint = {astro-ph/0301367},
 primaryClass = {astro-ph},
       adsurl = {https://ui.adsabs.harvard.edu/abs/2003ApJ...586L..19M}
}

@article{Waters2019Non-isobaricInstability,
    title = {{Non-isobaric Thermal Instability}},
    year = {2019},
    journal = {The Astrophysical Journal},
    author = {Waters, Tim and Proga, Daniel},
    number = {2},
    pages = {158},
    volume = {875},
    publisher = {IOP Publishing},
    url = {http://dx.doi.org/10.3847/1538-4357/ab10e1},
    doi = {10.3847/1538-4357/ab10e1},
    issn = {1538-4357},
    arxivId = {1810.11500}
}

@ARTICLE{2007ApJS_mignone,
       author = {{Mignone}, A. and {Bodo}, G. and {Massaglia}, S. and {Matsakos}, T. and {Tesileanu}, O. and {Zanni}, C. and {Ferrari}, A.},
        title = "{PLUTO: A Numerical Code for Computational Astrophysics}",
      journal = {\apjs},
         year = 2007,
        month = may,
       volume = {170},
       number = {1},
        pages = {228-242},
          doi = {10.1086/513316},
archivePrefix = {arXiv},
       eprint = {astro-ph/0701854},
 primaryClass = {astro-ph},
       adsurl = {https://ui.adsabs.harvard.edu/abs/2007ApJS..170..228M}
}

@misc{Davies26,
author = {Davies, R. and Choudhury, P. P. and Bott, A. F. A.},
title = {{\it in preparation}},
year = {2026}
}

@article{lopez_collisional_2025,
	title = {Collisional whistler instability and electron temperature staircase in inhomogeneous plasma},
	volume = {91},
	issn = {0022-3778, 1469-7807},
	url = {https://www.cambridge.org/core/product/identifier/S0022377825000078/type/journal_article},
	doi = {10.1017/S0022377825000078},
	language = {en},
	number = {2},
	urldate = {2025-08-08},
	journal = {Journal of Plasma Physics},
	author = {Lopez, N.A. and Bott, A.F.A. and Schekochihin, A.A.},
	month = apr,
	year = {2025},
	pages = {E45},
}

@article{bott_kinetic_2024,
	title = {Kinetic stability of {Chapman}–{Enskog} plasmas},
	volume = {90},
	issn = {0022-3778, 1469-7807},
	url = {https://www.cambridge.org/core/product/identifier/S0022377824000308/type/journal_article},
	doi = {10.1017/S0022377824000308},
	language = {en},
	number = {2},
	urldate = {2024-11-28},
	journal = {Journal of Plasma Physics},
	author = {Bott, Archie F.A. and Cowley, S.C. and Schekochihin, A.A.},
	month = apr,
	year = {2024},
	pages = {975900207},
}

@article{meinecke_strong_2022,
	title = {Strong suppression of heat conduction in a laboratory replica of galaxy-cluster turbulent plasmas},
	volume = {8},
	issn = {2375-2548},
	url = {https://www.science.org/doi/10.1126/sciadv.abj6799},
	doi = {10.1126/sciadv.abj6799},
	language = {en},
	number = {10},
	urldate = {2024-11-28},
	journal = {Science Advances},
	author = {Meinecke, Jena and Tzeferacos, Petros and Ross, James S. and Bott, Archie F. A. and Feister, Scott and Park, Hye-Sook and Bell, Anthony R. and Blandford, Roger and Berger, Richard L. and Bingham, Robert and Casner, Alexis and Chen, Laura E. and Foster, John and Froula, Dustin H. and Goyon, Clement and Kalantar, Daniel and Koenig, Michel and Lahmann, Brandon and Li, Chikang and Lu, Yingchao and Palmer, Charlotte A. J. and Petrasso, Richard D. and Poole, Hannah and Remington, Bruce and Reville, Brian and Reyes, Adam and Rigby, Alexandra and Ryu, Dongsu and Swadling, George and Zylstra, Alex and Miniati, Francesco and Sarkar, Subir and Schekochihin, Alexander A. and Lamb, Donald Q. and Gregori, Gianluca},
	month = {Mar},
	year = {2022},
	pages = {eabj6799},
}

@article{vincent_design_2026,
	title = {Design of experiments characterising heat conduction in magnetised, weakly collisional plasma},
	copyright = {https://creativecommons.org/licenses/by/4.0/},
	issn = {2095-4719, 2052-3289},
	url = {https://www.cambridge.org/core/product/identifier/S2095471926101522/type/journal_article},
	doi = {10.1017/hpl.2026.10152},
	language = {en},
	urldate = {2026-05-14},
	journal = {High Power Laser Science and Engineering},
	author = {Vincent, T. A. and Ariyathilaka, P. and Creaser, L. and Danson, C. and Lamb, D. and Meinecke, J. and Palmer, C.A.J. and Pitt, S. and Poole, H. and Spindloe, C. and Thomas, P. and Tubman, E. and Wilson, L. and Garbett, W. and Gregori, G. and Tzeferacos, P. and Hodge, T. and Bott, A. F. A.},
	month = may,
	year = {2026},
	pages = {1--31},
}

@article{Mohapatra2019TurbulenceThermodynamics,
    title = {{Turbulence in the intracluster medium: Simulations, observables, and thermodynamics}},
    year = {2019},
    journal = {Monthly Notices of the Royal Astronomical Society},
    author = {Mohapatra, Rajsekhar and Sharma, Prateek},
    number = {4},
    pages = {4881--4896},
    volume = {484},
    doi = {10.1093/mnras/stz328},
    issn = {13652966},
    arxivId = {1810.00018}
}

@ARTICLE{2012MNRAS_mccourt,
       author = {{McCourt}, Michael and {Sharma}, Prateek and {Quataert}, Eliot and {Parrish}, Ian J.},
        title = "{Thermal instability in gravitationally stratified plasmas: implications for multiphase structure in clusters and galaxy haloes}",
      journal = {\mnras},
         year = 2012,
        month = feb,
       volume = {419},
       number = {4},
        pages = {3319-3337},
          doi = {10.1111/j.1365-2966.2011.19972.x},
archivePrefix = {arXiv},
       eprint = {1105.2563},
 primaryClass = {astro-ph.CO},
       adsurl = {https://ui.adsabs.harvard.edu/abs/2012MNRAS.419.3319M}
}

@ARTICLE{2012MNRAS_sharma,
       author = {{Sharma}, Prateek and {McCourt}, Michael and {Quataert}, Eliot and {Parrish}, Ian J.},
        title = "{Thermal instability and the feedback regulation of hot haloes in clusters, groups and galaxies}",
      journal = {\mnras},
         year = 2012,
        month = mar,
       volume = {420},
       number = {4},
        pages = {3174-3194},
          doi = {10.1111/j.1365-2966.2011.20246.x},
archivePrefix = {arXiv},
       eprint = {1106.4816},
 primaryClass = {astro-ph.CO},
       adsurl = {https://ui.adsabs.harvard.edu/abs/2012MNRAS.420.3174S}
}

@ARTICLE{choudhury,
    
AUTHOR={Choudhury, Prakriti Pal },
           
TITLE={Formation of multiphase plasma in galactic haloes and an analogy to solar plasma},
          
JOURNAL={Frontiers in Astronomy and Space Sciences},
          
VOLUME={Volume 10 - 2023},
  
YEAR={2023},
  
URL={https://www.frontiersin.org/journals/astronomy-and-space-sciences/articles/10.3389/fspas.2023.1155865},
  
DOI={10.3389/fspas.2023.1155865},
  
ISSN={2296-987X}}

@ARTICLE{2025MNRAS_wibking,
       author = {{Wibking}, Benjamin D. and {Voit}, G. Mark and {O'Shea}, Brian W.},
        title = "{Precipitation plausible: magnetized thermal instability in the intracluster medium}",
      journal = {\mnras},
         year = 2025,
        month = dec,
       volume = {544},
       number = {2},
        pages = {2577-2585},
          doi = {10.1093/mnras/staf1801},
archivePrefix = {arXiv},
       eprint = {2506.10277},
 primaryClass = {astro-ph.GA},
       adsurl = {https://ui.adsabs.harvard.edu/abs/2025MNRAS.544.2577W}
}

@article{Komarov_2016,
    author = {Komarov, S. V. and Churazov, E. M. and Kunz, M. W. and Schekochihin, A. A.},
    title = {Thermal conduction in a mirror-unstable plasma},
    journal = {Monthly Notices of the Royal Astronomical Society},
    volume = {460},
    number = {1},
    pages = {467-477},
    year = {2016},
    month = {07},
    issn = {0035-8711},
    doi = {10.1093/mnras/stw963},
    url = {https://doi.org/10.1093/mnras/stw963},
    eprint = {https://academic.oup.com/mnras/article-pdf/460/1/467/8115658/stw963.pdf},
}
\bibliographystyle{aasjournalv7}



\end{document}